# Light Element Abundances From z=0 To z=5

F. X. Timmes[1,2], J. W. Truran[2], J. T. Lauroesch[2,3,4] & D. G. York[2]

[1]Department of Physics and Astronomy
Clemson University
Clemson, SC   29634

[2]Department of Astronomy and Astrophysics
and
Enrico Fermi Institute
University of Chicago
Chicago, IL   60637

[3]NASA/Goddard Space Flight Center
Code 631
Greenbelt, MD   60637

[4]National Research Council Resident Research Associate

email:  fxt@burn.uchicago.edu
truran@nova.uchicago.edu
jtl@banzao.gsfc.nasa.gov
don@oddjob.uchicago.edu.

# ABSTRACT

Plausible ratios of deuterium to hydrogen D/H as a function of metallicity, time, and redshift are investigated. Guided by the heavy element abundance patterns observed locally in Galactic dwarf stars and at large redshift in quasi-stellar object absorption line systems, empirical evolution of the relative abundance ratios Li/D, B/D, N/D, O/D, and F/D for QSO absorption line systems are given for the possible evolutionary patterns in D/H. Limits imposed from the few observations of deuterium and hydrogen in QSO absorption line systems (QSOALS) at $z \sim 3$ are consistent with the expected amount of oxygen produced in stars for the amount of deuterium burned in stars.

The atomic properties suggest that the O/D and N/D ratios will be free of substantial ionization, continuum placement, or optical depth errors. The unique advantages of using the O/D and N/D ratio, instead of the D/H ratio, in determining the primordial abundance and subsequent evolution of deuterium is as an important and recurring concept of this paper. Transitions of O I and N I that may be useful in determining precise O/D and N/D ratios in QSOALS, along with some QSOALS spectra of systems that may be of use in subsequent follow-up searches, are given in the appendices.

Definitive detection of lithium, boron and fluorine in QSOALS at an appropriate redshift, may be able to assist in identifying their chief nucleosynthetic origin site(s), although the practicality of making positive detections is much lower than for O/D and N/D. Positive detection of *any* fluorine at a sufficiently large redshift ($z \gtrsim 1.5$) would strongly suggest a positive detection of the neutrino process operating in massive stars. Observation of these light elements at several redshifts could also provide a useful diagnostic on the conditions of the early universe.

Subject headings: cosmology: observational – cosmology: theory – galaxies: abundances – galaxies: evolution – quasars: absorption line

# 1. INTRODUCTION

Abundance determinations in QSO absorption line systems (henceforth QSOALS), and the dependence of these absorption system abundances on redshift, are beginning to illuminate the nucleosynthetic pathway taken by the gas in QSOALS. Between z ∼ 2 to 5[1] the QSOALS easily outnumber all other detectable tracers of cosmic structure and nucleosynthetic processes. The absorption line systems sample a broad range of redshifts, baryonic densities, temperatures, and ionization stages. They are numerous enough to yield well-defined statistical properties of these physical quantities. At a large enough redshift, the QSOALS may be primitive enough to retain memory of the conditions present at the onset of galaxy formation in the early Universe.

Several independent lines of observational evidence suggest that the QSOALS are high-redshift progenitors of galaxies like our own (e.g., York et al. 1986; Yanny & York 1992; Wolfe 1995; Lauroesch et al. 1996; Lu, Sargent, & Barlow 1996). In addition, several attempts have been made to describe the formation and evolution of QSOALS as arising from gravitational instabilities and radiative cooling. These numerical simulations suggest that the Ly$\alpha$ systems can develop naturally, in a hierarchical theory of structure formation dominated by cold dark matter bathed in photoionizing background. The high column density lines of neutral hydrogen, N(H I) > $10^{17}$ cm$^{-2}$, arise, in general, from radiatively cooled gas associated with collapsed regions undergoing galaxy formation (e.g., Hernquist et al. 1996; Katz et al. 1996). Low column density absorption lines, N(H I) ∼ $10^{14}$ cm$^{-2}$, are produced by an assortment of length scales in various stages of gravitational collapse (e.g., Katz et al. 1996; Hernquist et al. 1996).

Elements heavier than carbon have such a small abundance at high redshift that only elements with (a) very strong resonance lines, (b) abundances greater than 1 part in $10^5$ of solar (by number), and (c) transitions in easily accessible wavelength regions, can be detected routinely in QSOALS (Meyer, Welty, & York 1989; Lauroesch et al. 1996). The elements possessing all these properties that have received the most attention in QSOALS are H, C, N, O, Mg, Si, and Fe. The dominant ionization stages of these elements in interstellar gas near 10,000°K are H I, N I, O I, C II, S II, Mg II, Si II, and Fe II (Trajero et al. 1996; York & Kinahan 1979). Except for C II, these ionization stages have numerous ($\gtrsim 6$) resonance lines above the Lyman limit with a wide range of f-values. The strongest lines in each case permit direct measurements of abundances as small as 0.001 $Z_\odot$ when N(H I) > $10^{19}$ cm$^{-2}$, or as small as $10^{-5}$ $Z_\odot$ when N(H I) > $10^{21}$ cm$^{-2}$. Less abundant elements such as Cr, Mn, Ni, and Zn have 2 or more strong resonance lines in regions of the spectra that can be observed routinely, but have abundances between 1 part in $10^6 - 10^8$ (by number). Hence, these elements are generally limited to use in systems in which abundances are above 0.01 $Z_\odot$. Where they can be detected, these elements are extremely useful for defining grain depletion of the gas and the nucleosynthetic behavior of iron peak nuclei, but for tracing the earliest stellar processing they may be useful only if very high signal-to-noise ratios can be achieved, pushing detection limits to below 100 $\mu$Å.

Simple cosmological transformations of the abundance levels and trends observed in the Milky Way's halo and disk stars can be used to predict any element ratio as a function of redshift (Timmes, Lauroesch, & Truran 1995a). Attempts to interpret the abundances of heavy elements in QSOALS within this picture lead to the conclusion that, for all reasonable choices of $\Omega$ in a $\Lambda=0$ cosmology (the transformations chosen ensure independence of the Hubble constant), the heavy element abundances observed in damped Ly$\alpha$ systems are consistent, over more than two orders of magnitude in abundance, with an $\sim 3\times10^9$ yr delay time between the start of the big bang and the formation of a thin disk. In simpler terms: to force a fit to the QSOALS abundance trends, a time delay in star formation of about $2 \sim 3$ Gyr seems necessary.

---

[1] Lower case z will be reserved for redshifts, upper case Z will be reserved for metallicities.

However, the scatter in the QSOALS heavy metal abundances and the narrow redshift window over which they have been observed could conspire to give the illusion that chemical enrichment began at these same redshifts (S. M. Fall 1995, private communication). In addition, there may be selection effects, such as dust obscuration or significant line of sight variations, that could substantially broaden the present redshift window (Fall & Pei 1995; Pei & Fall 1995). Nor is it clear that observational trends of solar vicinity stars are the best ones to use in the transformations suggested by Timmes et al. (1995a) – perhaps an average over radial or vertical distance would be more appropriate. While choosing a cosmology is presently ambiguous, there is in fact only one *actual* time-redshift relation and not a spread. On the other hand, the age-metallicity relationship used in the transformations is fraught with (dominated by) scatter, some of which is real (if not all) and thus contributes an intrinsic uncertainty of the method. If these objections turn out to be the case, then the Timmes et al. conclusion about a few Gyr hiatus before abundance evolution commences may have to be abandoned or at least modified. Here we adopt the time delay concept as a heuristic framework within which to proceed.

The abundance of primordial deuterium plays an important role in cosmology and in considerations of the nature of the dark matter. This abundance permits a crisp determination of the baryonic-to-critical density ratio $\Omega_B$ of the universe (e.g. Copi, Schramm & Turner 1995; Fields 1996). Unfortunately, the precise value of the primordial deuterium abundance as determined by recent measurements of D/H [2] in QSOALS indicates variations of at least a factor of 5 at z $\sim$ 3. Stars burn their initial deuterium to $^3$He while they are still on the pre-main sequence, and the net decrease in the deuterium content in the universe is related directly to the net increase in metallicity from stellar evolution. The scatter present in the observed abundances of silicon and iron in QSOALS with redshifts between 2$\sim$3, suggests immediately that the D/H ratio may also be variable in this redshift regime. Depletion of deuterium may even show up before metal enrichment, depending on the Hayashi track timescales and total mass ejected by prestellar winds. Conventional chemical evolution models, however, indicate that deuterium is depleted by a factor of 2$\sim$4 from its primordial abundance to formation of the Sun to the present interstellar medium. (Truran & Cameron 1971; Copi et al. 1995; Fields 1996).

Since the primordial abundance and subsequent evolution of deuterium appear to be somewhat uncertain, in this paper we consider several plausible and schematic histories of the D/H ratio as a function of metallicity. Our schematic histories are chosen specifically to span the range of "primordial" D/H ratios reflected in current observational determinations. We do not attempt to justify the rates of astration implied by these histories. These schematic D/H evolutions are then transformed into time and redshift spaces according to the prescription given in Timmes et al. (1995a). Thus, as a function of redshift, predictions for the the lithium to deuterium, Li/D; boron to deuterium, B/D; nitrogen to deuterium, N/D; oxygen to deuterium, O/D; and fluorine to deuterium, F/D, ratios are derived. Attention is focused here on the histories implied by standard cosmological parameters. However, our exploratory calculations suggest that models with a small positive cosmological constant effectively allow a longer amount of time to pass in the apparently critical two to four redshift interval.

---

[2] A linear abundance ratio by number will be denoted as X/Y, for species X and Y. A logarithmic abundance ratio relative to solar abundances will be denoted [X/Y] = $\log_{10}$ X/Y - $\log_{10}$ (X/Y)$_\odot$.

In §4 we show that at least one of the above ratios, O/D, can be determined more reliably from an empirical viewpoint than the D/H ratio. The O/D ratio is sensitive to both astration of deuterium and to production of $\alpha$-chain nuclei. Since these two processes are physically anti-correlated, ratios of the two isotopes amplify the deuterium destruction – $\alpha$-chain enhancement effect. The other ratios hold special significance also. The N/D ratio is relevant to the question of whether nitrogen is produced as a primary or secondary element. We note that Al II could be of significance in the context of tracking elements produced secondarily by neutronization. However, Al II is not discussed in this paper, since it is not yet sufficiently well observed and its ionization state is not tied closely to deuterium, which complicates the interpretation.

While we do not address in detail whether the very different D/H values quoted for QSOALS are consistent with the expected amount of deuterium astration and the ensuing metal enrichment by supernovae, we do point out several independent pieces of evidence that indicate that the two (deuterium destruction and metallicity production) are commensurate and within model expectations.

Whether the evolution of elemental lithium and boron from primordial nucleosynthesis is driven mainly by cosmic ray spallation of CNO nuclei, classical novae, asymptotic giant branch stars, or by the neutrino process in massive stars is presently controversial (Timmes, Woosley, & Weaver 1995b; Vangioni-Flam et al. 1996). Measurement of Li/D and B/D abundances in QSOALS is discussed in §5 and may assist in identifying the nucleosynthetic site of origin of these elements; however, neither element is likely to be detected in QSOALS without extreme effort. Another ratio that potentially has great power in illuminating early nucleosynthesis is the F/D ratio. In fact, positive detection of any fluorine at a sufficiently large redshift ($z \gtrsim 1.5$) would suggest strongly a positive detection of the neutrino processes operating in massive stars (Snow & York 1982; Woosley et al. 1990; Timmes et al. 1995b). Fluorine may also be produced through neutron addition in $^{13}$C rich pockets during thermal pulsations on the asymptotic giant branch in intermediate− and low−mass stars (Jorissen, Smith, & Lambert 1992; Forestini et al. 1992). Thus, measurement of the F/D ratio at several redshifts can place constraints on the fraction of fluorine produced by massive stars and the fraction produced in stars of intermediate− and low−mass.

## 2. D/H AS A FUNCTION OF METALLICITY, TIME, AND REDSHIFT

Figure 1a shows the schematic histories of deuterium versus [Fe/H] considered here. These various, plausible evolutions constitute the basic input. The evolution labeled "A" is an extreme case and begins with D/H $\sim 2\times 10^{-4}$, a value set by the upper limit of the D/H ratio observed in the z=3.32 absorption system toward the QSO Q0014+813 (Chaffee et al. 1985, 1986, Songalia et al. 1994; Carswell et al. 1994; Rugers & Hogan 1996a). Values of D/H approaching $\sim 10^{-4}$ are also commonly cited as the primordial D/H, derived by applying astration corrections to local D/H values (e.g., Copi et al. 1995; Fields 1996). The large initial abundance is taken to decrease linearly with [Fe/H] until D/H = $4\times 10^{-5}$ at [Fe/H] = 0.0; the value typical of the interstellar medium at a distance and age appropriate to the birth of the Sun. The astration of deuterium then continues in a linear fashion until the present epoch is reached; i.e., D/H = $2\times 10^{-5}$ at [Fe/H] = 0.2 (e.g., York & Rogerson 1976; Linsky et al. 1995). The curve labeled "B" starts with a smaller primordial deuterium abundance, D/H = $8\times 10^{-5}$, but it follows the same pathway with [Fe/H] as case "A". The evolution labeled "C" starts with the presolar abundance ratio of D/H = $4 \times 10^{-5}$, remaining constant with [Fe/H] (which requires some infall of fresh primordial material) until [Fe/H] = 0, at which point its evolution is the same as case "A". The curve labeled "D" is an extreme case that balances the destruction of deuterium with infall of fresh primordial material such that the abundance ratio D/H is the local value and its history is independent of metallicity. These four simple models contain the main components of abundance evolution with varying degrees of astration and, despite their simplicity, they place bounds on what is to be expected observationally.

The age-metallicity (time vs. [Fe/H]) relationship inferred from spectral observations of dwarf stars in the solar vicinity is shown in the inset plot of Figure 1b (Wheeler, Sneden, & Truran 1989; Edvardsson et al. 1993; Timmes et al. 1995b). The time axis in the inset plot is scaled between 0 and 1 by dividing by the age of the Universe, whatever value that may be for a chosen cosmology. Transformation of the metallicity axis in Fig. 1a by the age-metallicity relationship yields the schematic D/H evolutions as a function of time in Fig. 1b. A value of 15 Gyr was used for the age of the galaxy, a value chosen simply to have a concrete example. Confluence of curves at [Fe/H] = 0 in Fig. 1a maps into the 10.5 Gyr time point of Fig. 1b.

Following Timmes et al. (1995a), one can continue onward and use a cosmological redshift-time relationship to transform the time axis in Fig. 1b into a redshift axis (see the objections to this procedure in §1). The resulting D/H evolutions as a function of redshift are shown in Figure 1c, for several choices of $\Omega$ in a $\Lambda = 0$, $\tau_{\rm delay} = 2$ Gyr cosmology. The sharp kinks at z $\sim$ 2.5 in a $\Omega$=1 cosmology, and at z $\sim$ 4.2 in a $\Omega$=0.2 universe, occur at (and because of) the onset of stellar deuterium destruction. Firm observation of such a kink (z $\sim$ 3.5 from the putative measurements shown in Fig. 1c) would provide a confirmation and a quantification of $\tau_{\rm delay}$.

Five observations are shown in Fig. 1c. The lowest determined D/H ratio, $2.3 \times 10^{-5}$, is for a two component absorption system at z=3.572 along the line of sight to Q1937−1009 (Tytler, Fan & Burles 1996). The uncertainties are $\pm 0.3 \times 10^{-5}$, $1\sigma$ for statistical and seperately for systematic errors. Determination of the neutral hydrogen column density is the greatest source of the uncertainty, which Tytler et al. estimated to be N(H I) $\simeq 8.7\times 10^{17}$ cm$^{-2}$. Both [Si/H] and [C/H] were determined to be $\leq 0.01$ Z$_\odot$. Absorption at z=3.32 along the line of sight to S5 0014+813 provides an observation of D/H = $19.0 \pm 4.0 \times 10^{-5}$ (Rugers & Hogan 1996a). A second absorption system toward S5 0014+813 at z=2.798 has been reported to have D/H = $19.0(+1.6, -0.9) \times 10^{-5}$ by Rugers & Hogan (1996b)

Two upper limits are also shown in Fig. 1c. Carswell et al. (1996) give a limit of D/H $< 2 \times 10^{-4}$ from several different arguments, for an absorption system at z=3.087. They present suggested detections of deuterium in two components at $-79$ km s$^{-1}$ and $-93$ km s$^{-1}$ with respect to a multicomponent H I system at that redshift in Q0420$-$358. For all components in the system, N(H I) $\simeq 2 \times 10^{19}$ cm$^{-2}$, spread over 400 km s$^{-1}$, while for the two components noted, the total N(H I) $\simeq 5 \times 10^{18}$ cm$^{-2}$. They note that the suggested deuterium lines may be overlapping Ly$\alpha$ lines from other redshifts, but many lines of O I are definitely detected (these lines are are noted in appendix B of this paper). They also find O/D > 2 in a system that they suggest has reached a metallicity 1/10 of Solar in O, Si, and Fe (also see Fig. 3). Wampler et al. (1996) give a limit of D/H $< 1.5 \times 10^{-4}$ in a simple system at z $\leq$ 4.672, with N(H I) $\simeq 5 \times 10^{16}$ cm$^{-2}$. We label this point in Fig. 1c with an X because the proximity of the QSO redshift ($z_{em}$=4.694) and a possible detection of oxygen (implying a super-solar value for O/H) both suggest that the gas may be associated directly with the QSO.

While subject to large uncertainties, these five measurements suggest that different D/H values can occur within a small redshift range, assuming that the present scatter in the measurements is real. While early Galactic D/H measurements allowed the possibility of local variations in D/H (Vidal-Madjar et al. 1977), the newer data combined with the old suggest a quite uniform Galactic value (Linsky et al. 1995). Fig. 1c suggests that between z=3 and the present, variable D/H at a given redshift may still be found. Any possible variability in D/H injected near z=3 will take some time to damp out through astration and mixing of the gas, since star formation from galaxy to galaxy will not in general be coeval. Furthermore, the small region actually sampled in the QSO beam may prohibit the necessary averaging over a entire galaxy to see a direct correlation between astration of deuterium and production of heavy elements if the mixing timescales are slow compared to the galaxy evolution timescales. The same amplitude of scatter should be reflected in abundances of the elements formed as D is burned, as is evident from the scatter in the Zn/H and Si/H ratios for QSOALS between z $\sim$ 2$\sim$3. (e.g., Figs. 2 and 3 of Timmes et al. 1995a). Trends in these ratios should be in opposite directions for Si/H and D/H, since they are physically anticorrelated (deuterium burns and heavy elements are produced). When reliable data on many elements become available, the scatter on an absorber to absorber basis should be anti-correlated for D and O, Si, and Fe. However, in the context of standard stellar evolution, supernova nucleosynthesis, and galactic chemical evolution models, the deuterium destruction that accompanies the onset of heavy element formation is not expected to occur rapidly enough to account for a rapid variation in D/H ratios over a small redshift interval.

We have adopted [Fe/H] = $-3.0$ as the smallest gas metallicity that needs to be addressed in the calculations of Fig. 1 (and throughout this paper). There is, of course, a quantitative difference if one accepts a smaller or larger base metallicity. The qualitative points, however, remain unchanged. J. P. Ostriker (1995, private communication) has pointed out that a generation of Population III stars, generating 0.1% of the stars now known and 0.1% of the heavy elements we see today, would leave a floor of metallicity of [Fe/H] = −3.0 (at least for $\alpha$-chain elements lighter than iron) from z ∼ 20 to the epoch of halo star formation. This type of Population III stars could explain the detection of C IV features in the stronger lines of the Ly$\alpha$ forest (Meyer & York 1987, Cowie et al. 1996), as well as the absence of a detected Gunn-Peterson effect (Gunn & Peterson 1965) out to the highest redshifts observed thus far (Songaila, Cowie, & Lilly 1990). Audouze & Silk (1995) suggest that there is a natural metallicity threshold of $10^{-4}$ $Z_\odot$ due to first generation Type II supernovae. It is possible that such Pop III stars may help to explain the distinctive abundance patterns found in the 33 stars with metallicity [Fe/H] < −3.0 in the McWilliam et al. (1995a, 1995b) survey (see also Ryan, Norris, & Beers 1996) and some metal-poor compact blue galaxies (Kunth et al. 1994; Pagel 1992, 1994). Our assumption of a [Fe/H] = −3.0 base gas metallicity does not imply that there are no stars with a smaller metallicity; such stars would just be extremely rare in the absence of a z∼20 Population III.

## 3. ON PRECISE DEUTERIUM MEASUREMENTS

Ironically, it appears that while variable D/H in our Galaxy would be an argument against the primordial origin of deuterium, variable D/H ratios in the QSOALS sample is a straightforward consequence of the primordial origin of deuterium. The availability of high precision D/H measurements would enable more reliable determinations of cosmological parameters, rather than just showing qualitatively that the deuterium abundance is larger at z > 3 than at z=0. However, attempts to measure D/H are difficult. There are three principal problems. First, there are a plethora of lines in QSOALS spectra and there is no clean method to separate the lines of D I that accompany the H I source, from intrinsically weaker sources of H I. Second, virtually all strong Ly$\alpha$ forest lines are blends of many components. This makes it difficult to study that part of the H I absorption profile in which the optical depth in any one component is ∼ 1. The use of weaker parts of the profile is subject to continuum uncertainties, while using stronger parts is subject to saturation errors. Third, the heavy element line absorption systems display complex velocity structures, somewhat similar to that observed in the local interstellar medium (Ferlet et al. 1980, Welty, Hobbs, & Morton 1996). It is difficult to obtain sufficiently high signal to noise ratios and resolution in QSOALS spectra to define the velocity structure.[3]

To give a concrete example of these points, a moderate signal-to-noise spectrum of the suspected deuterium features of the z=3.32 absorption system toward QSO S5 0014+813 is shown in Figure 2a, and a D/H spectrum in the direction of a nearby star, $\zeta$ Puppis (Vidal-Madjar et al. 1977), is shown in Figure 2b. While the signal-to-noise ratio is evidently very high for the $\zeta$ Puppis, the D/H ratio derived is uncertain by at least $\pm$ 50% in these *Copernicus* satellite spectra (York & Rogerson 1976; Linsky et al. 1995). The deuterium feature in $\zeta$ Puppis requires fitting of an unknown feature (marked with a "U" in Figs. 2a and 2b). The components D1 and D2 are inferred, since they are obviously not resolvable in either deuterium or hydrogen. Furthermore, Ly$\alpha$ and Ly$\beta$ lines give column densities of neutral hydrogen that differ by a factor of 2 (York & Rogerson 1976; Morton 1978; Linsky et al. 1995). Fig. 2b shows that component blending problems are likely to exist for QSOALS spectra. The complication of neutral hydrogen blends will limit accuracy of N(H I) determinations associated with the alleged deuterium components. Weak H I lines in the Ly$\alpha$ forest that blend with the D I features will limit undisputed detections of D I to a small number. The complexity of QSO spectra leads to the expectation that abundance determinations as accurate as those obtained from interstellar spectra will be very difficult. Fig. 2b emphasizes, as an example, that values toward Galactic stars are still somewhat uncertain on a case by case basis.

---

[3] For studies of the local interstellar medium, there are particular advantages to using Na I, Ca II, K I and Ti II features for determining velocity structure on scales of 0.5 km s$^{-1}$. This information can be used to interpret the strong lines of more abundant elements observable only at resolutions < 4 km s$^{-1}$ and limited by the size of the instruments that can be launched into space for studies of the resonance lines of the abundant elements, all of which are in the ultraviolet. Unfortunately, Na I, Ca II, K I and Ti II are not routinely detected in QSOALS, although Robertson et al. (1988) and Meyer & York (1992) present some exceptions.

Rugers & Hogan (1996a) suggest that D/H pairs along the line of sight to S5 0014+813 clearly show a pattern b(D) / b(H) $\sim$ 1/1.4, as expected for thermally broadened lines (where the line width parameter b = $(2kT/m)^{1/2}$ km s$^{-1}$) at T $\sim$ 10,000 K. However, the errors are large enough that the ratio of the b-values could differ by a factor of 2. The average separation of the neutral hydrogen components is about 80 km s$^{-1}$, similar to the intrinsic separation of the D and H Ly$\alpha$ lines. Since any single H I component velocity is uncertain by 5 – 10 km s$^{-1}$, due to unnoted blends and signal-to-noise ratio limits, the selection of D components cannot depend on the velocity separation alone. For this reason, the selection of the two weak deuterium components was not based on the velocity separation (of which there are many possible D and H pairs), but on the fact that these two are the lowest equivalent width lines noted. This approach is reasonable and we agree with their analysis. However, basing the selection on finding low equivalent width lines, in general, may prejudice the ratio D/H toward small values of D/H (which could also represent processed material). Thus, the measurement of the "true" D/H ratio will require a large number of measurements of possible H and D pairs that show both 1.0 $\leq$ b(H) / b(D) $\leq$ 1.4 and $\Delta$ v $\sim$ 82 km s$^{-1}$. It is critical to the fitting process to have an exact velocity, based either on positions of neutral heavy element absorption lines (e.g., N I and O I), or several members of the deuterium and hydrogen Lyman series. The D/H patterns of Fig. 1c should then agree with the data.

Searches for ideal QSOALS in which to observe deuterium will continue. In this regard, Appendix A gives profiles of some deuterium and silicon features in QSOALS that were observed during the course of thesis research of one of us (JTL), and this may be of use in subsequent follow-up searches.

### 4. ARGUMENTS IN FAVOR OF USING OXYGEN

Many of the difficulties in obtaining a high precision deuterium abundance determination would be mitigated by measuring the O/D and N/D ratios. The H I, D I, N I, and O I features share three important characteristics of atomic physics. First, column densities of the neutral species dominate over column densities of ionized species in the absence of an intense stellar UV radiation field. Significant corrections for ionization are therefore not required to infer abundances from observations of neutral species. Second, the ionization potentials of these neutral lines differ by less than 1 eV. Thus, these species are locked by charge-exchange reactions to H I (e.g., $D^+ + H \leftrightarrow D + H^+$), and measured abundance ratios should give the true abundance ratios of the gas. Third, these features have several ultraviolet resonance transitions ($\gtrsim 6$) over a wide range of f-values (1 to 0.0001). Since there are multiple lines with a range of oscillator strengths, some feature can almost always be found near the same part of the curve of growth as for a line of the comparison species (York & Kinahan 1979). This fact essentially removes systematic errors due to a lack of knowledge of the proper component models.

There are other more practical considerations that encourage the use of these other species. Relative b-values of H I, D I, N I, and O I have been used to detect the warm neutral gas toward $\alpha$ Virginius and $\lambda$ Scorpii (York & Kinahan 1979; York 1983). Measurements of b(O I) and b(N I) in QSOALS, along with those of deuterium and hydrogen, would place the b-value arguments of Rugers & Hogan (1996a) on firmer ground. The narrow lines of D I, N I, and O I are also much less subject to continuum definition problems than the broader lines of H I. Furthermore, the N I and O I features can be observed in QSOALS even when the total metallicity is $< 10^{-3}$ $Z_\odot$.[4] Finally, the lines of H I, D I, N I, and O I are well separated, with many of the N I and O I lines occurring in triplets, so that complete obliteration of most of the lines of a species by accidental overlap from an unrelated Ly$\alpha$ forest line is extremely unlikely (Kulkarni et al. 1996).

In summary, the combination of atomic properties shared by H I, D I, N I, and O I (neutral species domination, charge-exchange locking to H I, and similar curve of growth locations), the atomic properties shared by D I, N I, and O I but not H I (narrow lines), and an exact velocity for the fitting process from N I and O I transitions (or several members of the deuterium and hydrogen Lyman series) suggest that concentrated studies of D I, N I and O I features will give empirical values of the O/D and N/D ratios that are free of substantial ionization, continuum placement, or optical depth errors. Thus, the O/D and N/D ratios, instead of the D/H ratio, offer unique advantages in determining the primordial abundance and subsequent evolution of deuterium.

Predictions of the O/D history with redshift are shown in Figure 3a, while the corresponding N/D histories are shown in Figure 3b. The curves were calculated with the cosmological models indicated on the plot in addition to a 3 Gyr starformation hiatus from the big bang. The inset plot in each figure represents a schematic evolution of the oxygen to iron ratio [O/Fe] (Fig. 3a) or the nitrogen to iron ratio (Fig. 3b) with metallicity [Fe/H], as inferred from surveys of solar neighborhood dwarf stars (e.g., Wheeler et al. 1989; Timmes et al. 1995b). The shaded region of the inset plot in Fig. 3b indicates roughly the amount of scatter in the abundance determinations. Stellar [Fe/H] values are converted into a time coordinate by the age-metallicity relationship of

---

[4] For systems with N(H I) $\gtrsim 10^{20}$ cm$^{-2}$; detectability is defined here as having at least 2 lines with expected equivalent widths $> 10$ mÅ. For systems with N(H I) $> 10^{17}$ cm$^{-2}$, the strongest transitions of N I and O I have expected equivalent widths $> 5$ mÅ for metallicities of 0.25 $Z_\odot$ and 008 $Z_\odot$, respectively.

Fig. 1b. Time coordinates are then mapped into redshift space by the redshift-time relationship of a chosen cosmology. Stellar [O/Fe] values are translated into [O/H] by dividing the appropriate [Fe/H] values of Fig. 1b. The [O/H] values are transformed into final O/D values by the assumed D/H histories of Fig. 1c (see objections and limitations of this procedure in §1). A similar prescription holds for nitrogen. It should be noted that Pettini, Lipman, & Hunstead (1995) have suggested that the N/O ratio in QSOALS may be significantly smaller in at least a fraction of (if not all) QSOALS than assumed above. If this were correct, our histories for O and N would have to be revised. Additional observations using several lines of N I and O I should help to confirm (or refute) the low N/O ratios observed by Pettini et al.

Elemental ratios of O/D and N/D in QSOALS, inferred from observations, are shown as boxed regions in the main plot of Fig. 3. The ranges of oxygen and nitrogen abundances, which determined the vertical dimension of the boxes, were deduced from observed Galactic halo star ratios in conjunction with the observed column densities of Si II and Fe II in QSOALS. Depletion of refractory elements onto dust grains has been ignored in making these estimates, although the required correction for dust can be made easily when the amount of depletion is significant. The horizontal dimensions of the boxes were set by the range of cosmological models considered. Deuterium abundances were obtained from the corresponding H I column densities (see Fig. 1c) for the H I values that correspond to minimum and maximum observed deficiencies of Fe and Si. The upper box assumed the smallest observed D/H ratio, $1.5 \times 10^{-5}$, while the lower box employed the largest observed D/H ratio of $2 \times 10^{-4}$. Downward pointing arrows on each box reinforce the notion that these regions are upper limits, since Fe II and Si II lines were used (see discussion above). The upper edge of the boxes shown in Fig. 3 are, in every sense, the envelopes of upper limits currently expected for O/D and N/D ratios in QSOALS. Note that the O/D > 2 finding by Carswell et al. (1996) fits inside the upper limit box. For other values of the delay time, the behavior of the curves can be inferred from Timmes et al. (1995a). For a 1 Gyr delay, the vertical portion of the curves for $\Omega=1$, $\Lambda=0$ lies near z=4.4. For the $\Omega=0.2$, $\Lambda=0$ case, the curves are still basically horizontal, with O/D=1 at z=5, that is, inconsistent with all the data based on which the boxes are drawn.

To assist in placing real observed values on Fig. 3, instead of simply inferred limit regions, Appendix B gives several features of the O I and N I lines that may be useful in determining precise abundances of these elements in QSOALS. This appendix serves as a useful update to the Morton (1991) compilation. Multiple lines are necessary in order to have a few that are not obliterated by chance Ly$\alpha$ forest lines (see Wampler et al. 1996), and in order to resolve saturation ambiguities implicit in interpreting the strongest, most easily detected features.

The arguments given above in favor of D/H versus O/D and N/D do not necessarily hold for species of other elements in H I regions, namely Si II and Fe II. These species are not locked to H I and D I by charge-exchange reactions but do exist in H II regions. Not only are the required ionization corrections for an isolated region uncertain, but line-of-sight H I and H II regions with the same or overlapping velocities cannot be disentangled. The amount of ionized hydrogen required to infer a total Si/H or Fe/H ratio is not directly measurable, thus all values of Si/H and Fe/H should be treated as upper limits. The effect is seen clearly in Fig. 4 of York & Kinahan (1979), which has an H I and H II region of comparable column density toward α Virginius. In at least two QSOALS , the z=1.78 absorber toward Q 1331+170 (Kulkarni et al. 1996) and the z=1.79 absorber toward B2 1225+317 (Lauroesch 1995), the derived N(SI II) / N(O I) ratio is 0.2∼0.3. For a solar metallicity H I region, the ratio should be $\simeq 0.03$, assuming no dust depletion of silicon. This suggests that corrections to Si/H in these systems are at least a factor of 10, in the direction of reducing Si/H (and Fe/H) abundances. The inclusion of H II region lines could lean in addition to inferences of b-values that are too large for the part of the Si II or Fe II feature that belongs with the H I and D I gas. Likewise, the positions of Fe II and Si II lines in QSOALS do not provide precise enough velocities for H I and D I profile fits. Use of Fe II and Si II lines for this purpose will most likely lead to incorrect velocities and b-values, due to blending of components from both H I and H II regions. In summary, reliance on the component structure of Si II and Fe II are likely to impede high precision deuterium abundance determinations.

On a related issue, zinc is used commonly as a metallicity standard for QSOALS as it is observed readily and believed to be free from depletion onto dust grains (Meyer et al. 1989; Pettini et al. 1994; Lu, Sargent & Barlow 1996). However, elemental zinc (chiefly the most abundant isotope $^{64}$Zn) has a complex nucleosynthetic history. It is probably made by three entirely distinct nucleosynthetic processes: (1) the weak component of the s-process in massive stars; (2) the alpha-rich freeze out in the inner most zones ejected (or not ejected) by the supernova during shock passage; and (3) the neutrino powered wind (Woosley & Weaver 1995; Hoffman et al. 1996). These same neutrino powered winds are required to make the r-process elements (e.g., osmium), which are observed to be primary (vs. secondary) nucleosynthetic products (Cowan et al. 1996; Sneden et al. 1996). If present supernovae calculations are correct, then this last source may be the largest contributor to the zinc abundances (Hoffman et al. 1996). The mass of zinc (and iron) ejected by Type II supernovae could even vary a great deal from star to star, due for example to the individuality of convection motions during core silicon burning. Standard Type Ia supernova models make about 1/2 the solar iron abundance but eject almost no zinc (Thielemann et al. 1986). Yet, observations of the zinc to iron ratio in Galactic dwarf stars suggests strongly that the ratio is flat and solar, at least for metallicities larger than 0.001 $Z_\odot$ (Sneden, Gratton, & Crocker 1991). The synthesis of zinc in supersolar proportions in some massive stars might then be required in order to explain the halo dwarf observations.

It would be better to have an element other than zinc as a metallicity indicator for QSOALS that can be interpreted without reference to dust depletion (as with zinc) but has a simpler nucleosynthetic history (unlike zinc). Oxygen may be the best choice. The chief reason for this preference is that oxygen is synthesized from a single, well-defined source: hydrostatic helium burning in presupernova stars. While oxygen abundances in QSOALS may be more difficult to measure, nonetheless, oxygen may well be a superior metallicity indicator.

## 5. OTHER LIGHT ELEMENTS

The primordial abundances of He, Li, Be, and B provide important information about the early universe. Evolution of lithium, beryllium, boron, and fluorine from their primordial abundances is presently controversial, as massive stars (especially the neutrino process), classical novae, asymptotic giant branch stars, Wolf-Rayet winds, and cosmic-ray spallation processes each have their proponents. Definitive detection of lithium, boron, and fluorine in QSOALS at an appropriate redshift may be able to assist in identifying their chief nucleosynthetic origin site(s), although the practicality of making positive detections is much lower than for O/D and N/D. While oxygen may do as well or even better as the the comparison element (e.g., Li/O), we will focus on using deuterium (e.g., Li/D) as the reference element since its deviation from its primordial abundance is expected to be small (see discussion in §1).

Measurement of He in QSOALS differ from the other elements advocated above in at least two respects. First, large ionization corrections are required to compare measured He I (and/or He II) column densities with hydrogen abundances. Second, the ground state resonance lines of He I and He II are at wavelengths shorter than the Lyman limit. This second difference means that one can observe these lines only in systems in which the total H I column density $\lesssim 4\times10^{17}$ cm$^{-2}$, in conjunction with there being no opaque Lyman limit systems at other redshifts. For example, QSO S5 0014+813 has 2 associated Lyman limit systems, one at z∼2.8 and an opaque limit at z∼2.4. In some cases, notably the QSO HS 1700+6416 (Reimers & Vogel 1993), He I absorption has been detected. As Reimers & Vogel show, however, the derived helium abundance depends sensitively on the assumed ionizing background.

Lithium is never seen in any state but the neutral state, and it is a trace species in all interstellar conditions. Thus, it is unlikely that lithium will ever be seen in QSOALS, due to the expected low column densities of Li I and lack of suitable lines. The one resonance line available is at 6707 Åand will be redshifted out of the range of low noise optical detectors in almost all circumstances. Nevertheless, the expected Li/D ratio as a function of redshift is shown in Figure 4. The small inset figure shows a schematic of the upper envelope of lithium abundances deduced from surveys of stars in the solar vicinity (see, for example, Timmes et al. 1995b). Transformation of the small inset plot into the main plot of Fig. 4 follows the previous prescriptions. At a sufficiently large redshift (z $\gtrsim$ 3), the lithium abundance in QSOALS should become a constant, reflecting the primordial value (Spite plateau). If significantly larger values of lithium are detected, these might constitute evidence in favor of inhomogeneous primordial nucleosynthesis. At smaller redshifts, the upper envelope of lithium observed in QSOALS should increase smoothly, up to a maximum value that is an order of magnitude greater than those found at larger redshifts. It is very easy to destroy lithium in a star by transport mechanisms (diffusive, convective, and meridional circulation) that move the lithium into regions in which the temperature is in excess of a few million degrees (Michaud & Charbonneau 1991). At these temperatures, lithium is burned quickly by (p,$\gamma$) and (p,$\alpha$) reactions. Hence, one may expect that below the upper envelope, lithium abundances in QSOALS will span the entire range of measurable abundances or have considerable scatter about a mean value.

There are two resonance lines of berylium at $\sim$ 3130 Å that are within the range of optical detectors for low redshift systems (z $\lesssim$ 2). However, abundance determinations of Be in QSOALS are hampered by the extermely small absolute abundance of Be. In addition, the relatively large condensation temperature of Be (e.g., Wasson 1985) suggests that significant corrections for depletion onto dust grains may be required. The redshift evolution of Be is expected to differ little from that of boron, if Galactic abundance trends of these two elements are indicative (Timmes et al. 1995a).

The ultraviolet B II resonance line near 1362 Å will be hard to detect, but if it can be measured the interpretation is more straightforward than for lithium since B II is a dominant ionization stage. Figure 5 shows the evolution of the B/D ratio with redshift. The small inset figure shows the schematic evolution of boron abundances with metallicity as inferred from surveys of stars in the solar vicinity. The linear increase with [Fe/H] suggests a primary origin of $^{11}$B (the dominant isotope). Transformation of the small inset plot into the main plot follows the previous mapping methods. The expected B/H evolutions as a function of redshift are shown for several choices of $\Omega$ in a $\Lambda = 0$, $\tau_{\rm delay} = 3$ Gyr cosmology. The N (H I) + N(H II) column densities for some Ly$\alpha$ systems may be $> 10^{22}$ cm$^{-2}$. Depending on the mechanism for boron nucleosynthesis, equivalent widths may then reach 10 mÅ. In such cases, correction for the ionization conditions (B II would also exist in H II regions) may be possible by inference from Si/O and similar ratios.

Fluorine is very difficult to measure in QSOALS, primarily because it is the least abundant of all the stable $12 \leq A \leq 38$ nuclides, and F I, the dominant ionization stage expected, has only two moderately strong lines above the Lyman limit. The ionization potential of F I (17 eV) is significantly larger than that of neutral hydrogen, so the observable species is, fortunately, the dominant ionization stage. The calculated history of the F/D ratio is shown in Figure 6. The small inset figure shows the schematic evolution of fluorine abundances with metallicity, as suggested by the very few observations that exist of the hydrogen fluoride molecule in solar neighborhood stars. At [Fe/H] $\lesssim -1.0$, the predicted history is guided by theoretical prejudices for the site of origin of fluorine (Timmes et al. 1995b). Mapping of the small inset plot into the main plot follows as before. Positive and undisputed detection of any fluorine at a sufficiently large redshift (z $\gtrsim$ 1.5) in QSOALS would constitute the strongest evidence to date for the existence of the neutrino process operating in massive stars (Snow & York 1981; Woosley et al. 1990; Timmes et al. 1995b; Vangioni-Flam et al. 1996). Intermediate- and low mass stars on the asymptotic giant branch can also produce fluorine (Jorissen at al. 1992; Forestini et al. 1992; Mowlavi, Jorissen, & Arnould 1996). In these cases, $^{13}$C produces neutrons via the $^{13}$C($\alpha$,n)$^{16}$O reaction. Some of these neutrons are captured by $^{14}$N to produce $^{14}$C and protons. These protons are then captured by $^{18}$O to produce fluorine by the sequence $^{18}$O(p,$\alpha$)$^{15}$N($\alpha$,$\gamma$)$^{19}$F. The fluorine is then dredged up to the surface following thermal pulsations. Observation of the F/D ratio at several redshifts would place interesting constraints on the fraction of fluorine due to massive stars and the fraction due to intermediate− and low−mass stars.

6. SUMMARY

Cosmic-chemical evolution may have remained at a threshold from the earliest times until z∼3 and subsequently led to the abundances we see today. This surprising scenario, in tandem with present and feasible observations of QSOALS abundances, offers the possibility for strong tests of Ω, as well of the nucleosynthesis of α-element formation (oxygen and silicon), elements beyond the iron peak (zinc), secondary neutronization (aluminum), and tentative models of lithium and boron production by cosmic rays, novae, or the neutrino process.

The calculated D/H evolutions with time, metallicity and redshift (Fig. 1) suggest that certain ranges for the primordial D/H ratio are consistent with recent D/H measurements in QSOALS. Sharp kinks occur in the redshift histories (Fig. 1c) due to the onset of deuterium destruction. Firm observation of such a kink (z ∼ 3.5 from the putative measurements shown in Fig. 1c) may be providing a confirmation and a quantification of $\tau_{\text{delay}}$. While we have not addressed in detail whether the very different D/H values quoted for QSOALS are consistent with the expected amount of deuterium astration and the ensuing metal enrichment by supernovae, we have pointed out several independent pieces of evidence (O/D >2 in at least one system, consistency with silicon, iron, and zinc abundances and their scatter) that indicate that the amount of deuterium destruction and the amount of metallicity production are compatible with model expectations.

Atomic physics suggests (see §3) that concentrated studies of D I, N I and O I features will give empirical values of O/D and N/D that are free of substantial ionization, continuum placement, optical depth and velocity errors. Abundances of these elements are predictable directly from simple transformation-based cosmological models and element trends observed in Galactic stars. The D/H ratio, and hence O/D and N/D, may vary erratically in the redshift region from 2 to 3, so it is important for observational cosmology to obtain measurements at z < 1.5 and at z > 3.5, in order to free the abundance comparisons from transient effects of galaxy formation. Interpretation of such observations will be more straightforward if the metallicities of QSOALS are based on the O/H ratios, instead of the currently preferred Zn/H ratio.

We thank the anonymous referee for several suggestions that added clarity and smoothness to this manuscript.

This work has been supported at Clemson by a Compton Gamma Ray Observatory Fellowship (FXT); at Chicago by NSF grant AST 92-17969 (JWT), NASA grant NAG 5-2770 (JWT), NASA subcontract NAS5-32985 to Johns Hopkins University, for the FUSE project (DGY), an Enrico Fermi Postdoctoral Fellowship (FXT); and at Goddard by a National Research Council Resident Research Associate grant (JTL). JTL thanks the Space Sciences Data Operations Office and the *Copernicus* Spectral Data Archive Project (in particular Michael Van Steenberg, Randy Thompson, and George Sonneborn) at NASA/Goddard Space Flight Center for their hospitality and support.

APPENDIX A: ADDITIONAL SEARCHES FOR MEASURABLE DEUTERIUM LINES

The measurement of D/H at high redshifts is limited to systems that meet criteria noted in the main text. A few ideal systems may emerge, but obtaining a large number of values will depend on evaluating numerous possible systems and deciding how to best use what is available in each case. Here we present a set of profiles for 3 systems toward the bright QSO HS 1946+7658 that may be useful in future searches. Table 1 and Figure 7 summarize the derived H I and selected other column densities. The three systems, in addition to our own data for S5 0014+813, are discussed briefly below.

The z=2.644 system toward HS 1946+7658 has a strong Ly$\alpha$ line with a rest equivalent width of $1.05 \pm 0.05$ Å, as well as narrow (b $\sim$ 9 km s$^{-1}$) C IV absorption. The strongest Si II line at 1260 Å is blended with a Ly$\alpha$ forest line, so no information on the less strongly ionized gas is available. If the center of the C IV lines is assumed to coincide with the main H I component, the profile fitting procedure requires the addition of a second H I component at the expected position of any D I feature that might be present. However, fitting of additional Lyman lines might be useful, particularly given the uncertain column density: $2 \times 10^{16}$ cm$^{-2} \leq $ N(H I) $\leq 9 \times 10^{17}$ cm$^{-2}$, with the smallest $\chi^2$ fit occurring when it is assumed that H I and C IV are coincident at $\sim 6.5 \times 10^{16}$ cm$^{-2}$. Sensitive searches for N I (1199 Å), O I (1302 Å), and Si II (1190, 1193 Å) are required (at observed wavelengths > 4000 Å) to direct the profile fitting.

The damped Ly$\alpha$ system at z=2.843 has N(H I) $= 1.6 \times 10^{20}$ cm$^{-2}$, and fitting of the deuterium features associated with Ly$\gamma$ through at least Ly$\epsilon$ could provide a very useful constraint on D/H. This system is useful as [O/H] $\sim -2.8$, so little astration of deuterium should have occured. However, there is a rather strong C IV component at $\sim -72$ km s$^{-1}$ that could have associated H I that may interfere with the detection of the D I lines (in a fashion similar to the "unknown" component toward $\zeta$ Pup in Fig. 2b). Sensitive observations of Si II at 1260/1190/1193 Å might reveal associated low ionization components at $-72$ km s$^{-1}$. If none is found, multiple components are not indicated. If Si II is found in z=2.843, a sensitive study of O I will be required to see if the components affect H I and the D/H fitting process.

The absorption system at z=2.893 shows a moderately strong Ly$\alpha$ line (rest equivalent width of $0.73 \pm 0.3$ Å) and 2 closely spaced C IV components, with the 120 km s$^{-1}$ component displaying Si III and Si IV absorption as well (see Fig. A1). Fits to the Ly$\alpha$ line for this system are uncertain; estimated upper limits to N(H I) can be made from the absence of a noticable Lyman limit. An extreme lower limit of $2 \times 10^{15}$ cm$^{-2}$ can be derived assuming the same H I components as measured for the metals, but with a b-value that is scaled by the elemental mass. The relatively strong metal lines suggest that this system may have a significant metal abundance, but fits to the other Lyman-series lines are needed to provide more stringent limits on the column density of H I gas. As with other systems, sensitive searches for O I (at $\lambda > 4000$ Å) are needed to guide the fitting process.

# APPENDIX B: LINES OF N I, O I and F I

Between 912 Å $< \lambda <$ 1356 Å there are over 15 lines for both N I and O I that span over 4 orders of magnitude in oscillator strength. The strongest of these lines begin to be detectable at optical wavelengths at redshifts of z $\sim$ 1.3, and by z $\sim$ 2.3 all of these lines are, in principle, detectable by ground-based telescopes. Morton (1991) presented atomic data for resonance absorption lines longward of the Lyman-limit for elements lighter than germanium. Since then, there have been some additional measurements and theoretical calculations of the oscillator strengths of lines of N I and O I, the results of which are summarized here. The wavelengths and oscillator strengths adopted by Morton (1991) are listed in Table 2, along with new results of selected calculations (e.g., Verner, Barthel, & Tytler al. 1994, Verner, Verner & Ferland 1996) and measurements. Also see the updates in Tripp, Lu & Savage (1996). Individual transitions are discussed below.

The species N I has numerous transitions below 1200 Å, including the strong multiplet at 1200 Å, as well as the somewhat weaker multiplets near 953, 964, and 1134 Å. As noted by Morton (1991), there is generally good agreement between laboratory measurements and theoretical calculation for the multiplets at 1200 and 1134 Å; the recent laboratory measurements of the 1134 Å multiplet by Goldbach et al. (1992) are only 0.5% and 13% larger than the values suggested by Morton. For the multiplets at 953 and 964 Å, there is some significant scatter in the f-values for the individual lines among laboratory, theoretical, and interstellar measures. For the 964 Å multiplet, the length values of Hibbert, Dufton, & Keenan (1985) are between 10% and 30% larger for the lines at 963.99 and 964.63 Å than the results of either Goldbach et al. (1992) or Hibbert et al. (1991a), or the values derived from the interstellar values of Lugger et al. (1978). There is good agreement, however, between all sources except Goldbach et al. (1992) for the line at 965.04 Å. For the multiplet at 953 Å, there is also some scatter among the various sources: the lines at 953.42 and 953.65 Å show good agreement between Lugger et al. (1978) and Hibbert et al. (1985, 1991a), while for 953.97 Å , the value of Hibbert et al. (1985) is somewhat smaller than the average of Lugger et al. (1978), Hibbert et al. (1991a), and Goldbach et al. (1992). Clearly there can be variations among the various methods, and further laboratory measurements, theoretical models, and interstellar measures (either from a reanalysis of *Copernicus* archival spectra and/or new *Far Ultraviolet Spectroscopic Explorer* observations) will be required to reduce the scatter in the measurements. In the absence of clear trends among the measures of oscillator strengths, the values of Hibbert et al. (1985) are suggested. The bottom panel of Figure 8 shows *Copernicus* observations of the N I transitions discussed above towards the star $\alpha$ Vir. These features are easily detected in Galactic lines-of-sight with low H I column densities, and should be detectable in damped Ly$\alpha$ systems with H I column densities $\gtrsim 10^{20}$ cm$^{-2}$ and metallicities $\lesssim 0.01$ Z$_\odot$.

For O I there are numerous multiplets below 1000 Å, as well as strong lines at 1036 and 1302 Å (the very weak intersystem line at 1355 Å will not be discussed here). The majority of the multiplets below 1000 Å are too weak to be observed in the majority of systems; however, the lines at 988.58, 988.66, and 988.77 Å are likely to be observable in many systems. As summarized by Zeippen, Seaton, & Morton (1977), a number of laboratory measurements have converged on an f-value of 0.4887 for the 1302 Å line; the recent calculations of Hibbert et al. (1991b) and Biémont & Zeippen (1992) both find an oscillator strength ∼7% larger than the value adopted by Zeippen et al. (1977) (and by Morton). Similarly, the f-value adopted by Zeippen et al. (1977) for the 1039.23 Å line is less than 5% different than the recent measurements of Goldbach & Nollez (1994) and the calculations of Hibbert et al. (1991b). For the strong multiplet at 988 Å, there are some significant variations among Zeippen et al. (1977), Hibbert et al. (1991b), Biémont & Zeippen (1992), and Goldbach & Nollez (1994). The very weak line at 988.58 Å, which was not measurable by Goldbach & Nollez, shows good agreement between Hibbert et al. (1991b) and Biémont & Zeippen, but both are ∼15% larger than that adopted by Zeippen et al. (1977). For the moderate strength line at 988.65 Å, the calculations of Hibbert et al. (1991b) and Biémont & Zeippen (1992) are again in good agreement, but again larger (∼13%) than Zeippen et al. (1977), while the measurement of Goldbach & Nollez (1994) is in turn ∼18% larger than the average of Hibbert et al. (1991b) and Biémont & Zeippen (1992). There is, however, excellent agreement between the four sources for the strongest (and most useful) line at 988.77 Å, with an average value of f = 0.0453 ± 0.003. The O I multiplet at 971 Å (UV multiplet 10) is also a reasonable candidate for detection in some QSOALS, having ≃ 30% of the total multiplet strength of the 988 Å multiplet, but with three lines separated by $\Delta\lambda = 0.0011$ Å (Zeippen et al. 1977). Unfortunately this multiplet was not included in the calculations of Biémont & Zeippen or the measurements of Goldbach & Nollez; the multiplet strength from Hibbert et al. (1991b) is ∼ 20% lower than that of Zeippen et al. (1977) (with little scatter among the lines), suggesting that the relative oscillator strengths for this multiplet are accurate but that there may be an overall shift of order 20%. The strong multiplet at 1025 Å will generally be masked by Ly$\beta$ and is not considered further here. The top panel of Figure 8 shows *Copernicus* observations of the O I transitions discussed above towards the star $\alpha$ Vir. As for nitrogen, these lines should be detectable in damped Ly$\alpha$ systems with H I column densities $\gtrsim 10^{20}$ cm$^{-2}$ and metallicities $\lesssim 0.01$ Z$_\odot$.

For neutral fluorine, there have been no recent measurements of the oscillator strengths of the F I lines at 951.87 and 954.83 Å. The results of Pinnington et al. (1976) differ by a factor of 2 from those of Clyne & Nip (1977); the oscillator strengths of Pinnington et al. are preferred, since the oscillator strengths for several C I lines analyzed by Clyne & Nip in the same paper are smaller by factors of ∼ 2∼4, for the measured lines, when compared to the Morton values. The Verner et al. (1994, 1996) compilations list oscillator strengths for the F I lines that are 91% of the Pinnington et al. values. However, if the Clyne & Nip oscillator strength is closer to the actual value, then this suggests that fluorine may be essentially undepleted in the Galactic interstellar medium. This effectively reduces the likelihood that F I may be detected in QSOALS, since for the same column density one obtainss a weaker line. We adopt the average of the Pinnington et al. and Verner et al. oscillator strengths here. Additional laboratory and theoretical work is called for, given the unique nucleosynthetic origins of fluorine and the likelihood that it will be detected in numerous lines of sight by the *Far Ultraviolet Spectroscopic Explorer* satellite.

## TABLE 1

Systems for New Searches for Deuterium Toward HS 1946+7658

| $z_{obs}$ | N(H I)[a] | N(D I) | N(O I) | N(Si II) |
|---|---|---|---|---|
| 2.664 | $6.5 \pm 1.8 \times 10^{16}$ | ... | ... | $< 4.8 \times 10^{12}$ |
| 2.843 | $1.6 \pm 0.1 \times 10^{20}$ | ... | $2.8 \pm 0.9 \times 10^{14}$ | $3.3 \pm 0.2 \times 10^{13}$ |
| 2.893 | $< 1.0 \times 10^{17}$ [b] | $< 1.4 \pm 0.3 \times 10^{13}$ | ... | $< 1.3 \times 10^{12}$ |

[a] All column densities in cm$^{-2}$  [b] Sum over both components

## TABLE 2

Atomic data for Selected Lines of N I, O I and F I for QSOALS

| Species | UV Multiplet | $\lambda$ (Å)[(1)] | Suggested f | Reference(s) |
|---|---|---|---|---|
| N I | 1 | 1200.7098 | 0.0442 | (1) |
|  |  | 1200.2233 | 0.0885 | (1) |
|  |  | 1199.5496 | 0.133 | (1) |
|  | 2 | 1134.9803 | 0.0403 | (1) |
|  |  | 1134.4149 | 0.0268 | (1) |
|  |  | 1134.1653 | 0.0134 | (1) |
|  | 3 | 965.0413 | 0.0058 | (2) |
|  |  | 964.6256 | 0.0118 | (2) |
|  |  | 963.9903 | 0.0184 | (2) |
|  | 3.05 | 953.9699 | 0.0259 | (2) |
|  |  | 953.6549 | 0.0203 | (2) |
|  |  | 953.4152 | 0.0106 | (2) |
| O I | 1 | 1302.1685 | 0.0489 | (1) |
|  | 3 | 1039.2304 | 0.0092 | (1),(3),(4) |
|  | 5 | 988.7734 | 0.0453 | (3),(4),(5),(6) |
|  |  | 988.6549 | 0.0087 | (3),(6) |
|  |  | 988.5778 | $5.9 \times 10^{-4}$ | (3),(6) |
|  | 10 | 971.7382 | 0.0112 | (3),(5) |
|  |  | 971.7376 | 0.0020 | (3),(5) |
|  |  | 971.7371 | $1.3 \times 10^{-4}$ | (3),(5) |
| F I | 1 | 954.826 | 0.0828 | (1),(7) |
|  |  | 951.870 | 0.0166 | (1),(7) |

(1) Morton 1991, (2) Hibbert, Dufton, & Keenan 1985, (3) Hibbert et al. 1991b, (4) Goldbach & Nollez 1994, (5) Zeippen, Seaton, & Morton 1977, (6) Biémont & Zeippen 1992 (7) Verner et al. 1994, 1996.

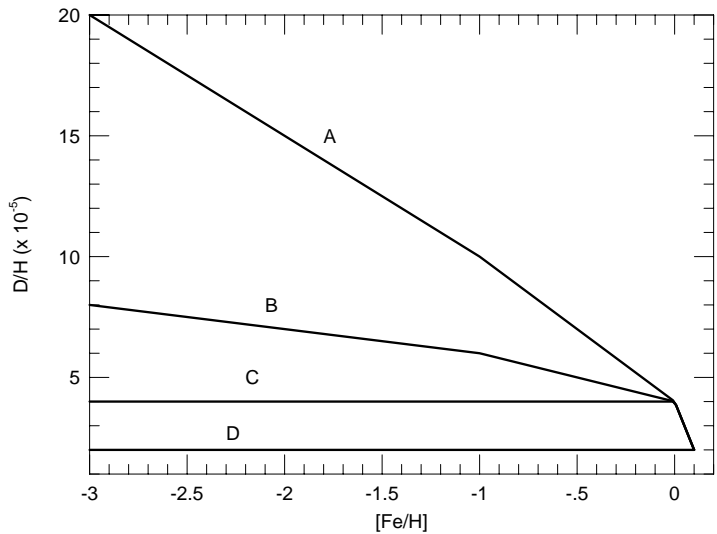

Fig. 1a.— Deuterium to hydrogen ratios, by number, as a function of [Fe/H]. These evolutions are schematic, represent various plausible evolutions, and are the basic input. Case "A" is begins with large value of D/H and decreases linearly with [Fe/H] until [Fe/H] = 0 (metallicity appropriate to the birth of the Sun), at which point the slope changes with a new linear decrease to the present interstellar medium. Curve "B" has the same basic evolution as "A", but it starts with a somewhat smaller primordial deuterium abundance. Evolution "C" starts with the presolar D/H and remains constant until [Fe/H] = 0, at which it then follows "A". Case "D" has a present epoch interstellar medium value and is independent of [Fe/H]. The latter two evolutions require some level of balance between deuterium destruction and infall of fresh primordial material.

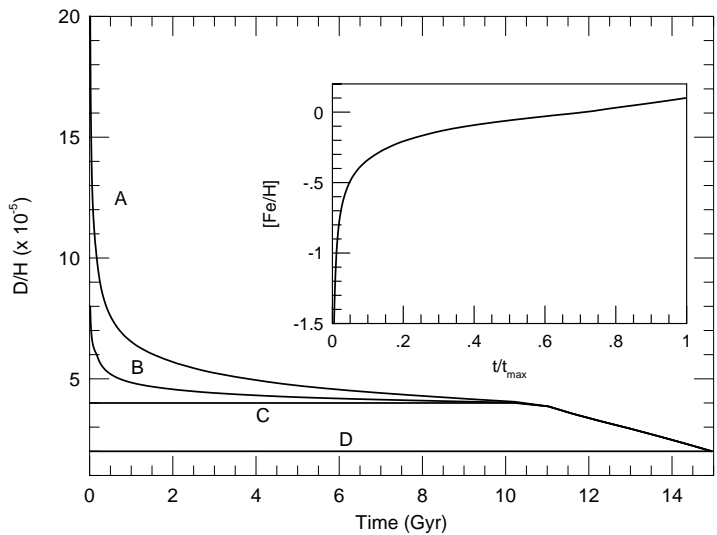

Fig. 1b.— Deuterium to hydrogen ratios, by number, as a function of time. The inset plot shows the (scatter free!) age-metallicity relationship deduced from spectral observations of nearby

dwarf stars. Its x-axis is scaled so that the evolution is independent of the assumed age of the universe. Transforming the metallicity axis of Fig. 1a by the age-metallicity relationship yields the main plot. A Galactic age of 15 Gyr was used in the main plot simply for illustrative purposes, with the age of the Sun accounting for the confluence of the curves at the presolar D/H=$4.0\times10^{-5}$, [Fe/H]=0.0 point. The labeled evolutions A–D are the same as in Fig. 1a.

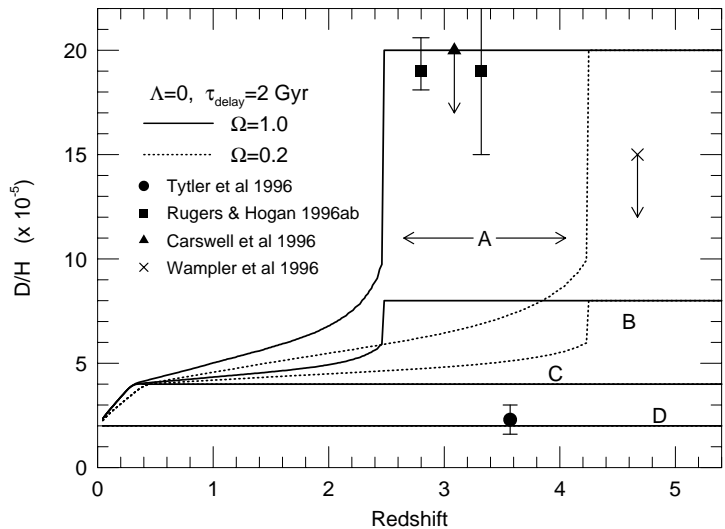

Fig. 1c.— Deuterium to hydrogen ratios, by number, as a function of redshift. A cosmological redshift-time relationship transforms the time axis of Fig. 1b into a redshift axis. The labeled evolutions A–D are the same as in Fig. 1a. Several choices of $\Omega$ in a $\Lambda = 0$, $\tau_{\text{delay}} = 2$ Gyr cosmology are shown. The sharp kinks at z $\sim$ 2.5 in a $\Omega$=1 cosmology, and at z $\sim$ 4.2 in a $\Omega$=0.2 universe, occur at (and because of) the onset of stellar deuterium destruction. Firm observation of such a kink (z $\sim$ 3.5 from the putative measurements shown here) would provide a confirmation and a quantification of $\tau_{\text{delay}}$. The lowest observed D/H ratio (filled circle), $2.3 \pm 0.7 \times 10^{-5}$, is along the line of sight to Q1937-1009 (Tytler et al. 1996). Two absorption systems along the line of sight to S5 0014+813 provide values of D/H = $19.0 \pm 4.0 \times 10^{-5}$ and D/H = $19.0(+1.6, -0.9) \times 10^{-5}$ (filled squares; Rugers & Hogan 1996a, 1996b). Carswell et al. (1996) give an upper limit of D/H $< 2\times10^{-4}$ for a system at z=3.087 (filled triangle). Wampler et al. (1996) give an upper limit of D/H $< 1.5 \times 10^{-4}$ for a system at z $\leq$ 4.672. This point is labeled with an X because the proximity of the QSO redshift ($z_{em}$=4.694), and a possible detection of oxygen, both suggest the gas may be associated directly with the QSO.

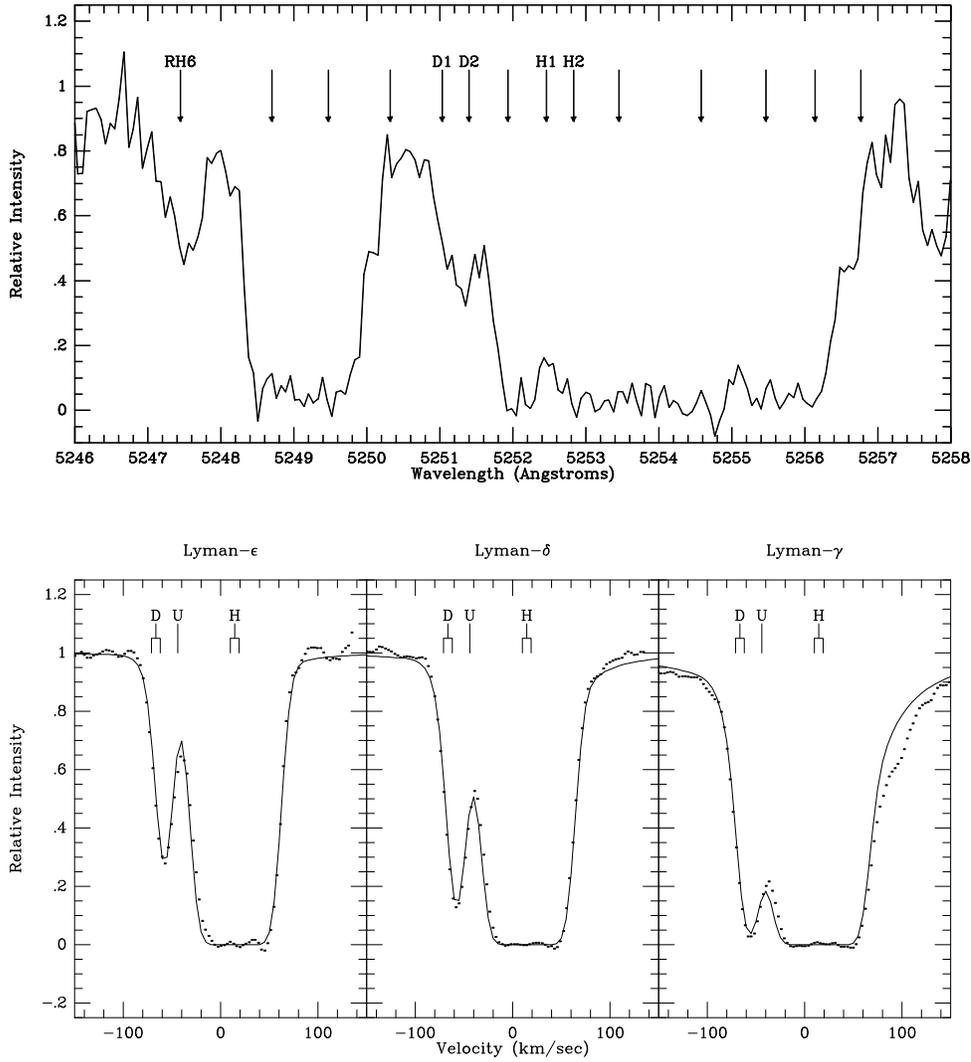

Fig. 2.— (a) Spectrum near Lyα of the z=3.32 absorption system toward QSO S5 0014+813 (KPNO 4-m, Lauroesch 1995). The wavelength scale is in the local standard of rest frame (e.g., Rugers & Hogan 1996a). Hydrogen lines from Rugers & Hogan are marked by ticks starting with component 6 (labeled RH6 in the figure). All features shown were shifted by 10 km s$^{-1}$ in order to align component 6 with the 5247.5 Å component 1 of Carswell et al. (1996). The Lauroesch (1995) and Carswell et al. wavelength scales agree to better than 3 km s$^{-1}$. The position of the two suspected deuterium features are labeled D1 and D2 (components 11 and 13 in Rugers & Hogan), with the corresponding hydrogen features labeled H1 and H2. Contrasting this spectrum with that of Rugers & Hogan emphasizes the need for very high signal-to-noise ratios. (b) *Copernicus* spectra of Lyγ, Lyδ, and Lyε lines in the direction of ζ Puppis. The fits to the spectra, using the parameters of Vidal-Madjar et al. (1977), are superimposed. Fits to these Lyman series spectra requires the addition of an unknown component, labeled "U", in addition to the 2 components $H_1/D_1$ and $H_2/D_2$ inferred from fitting the deuterium lines.

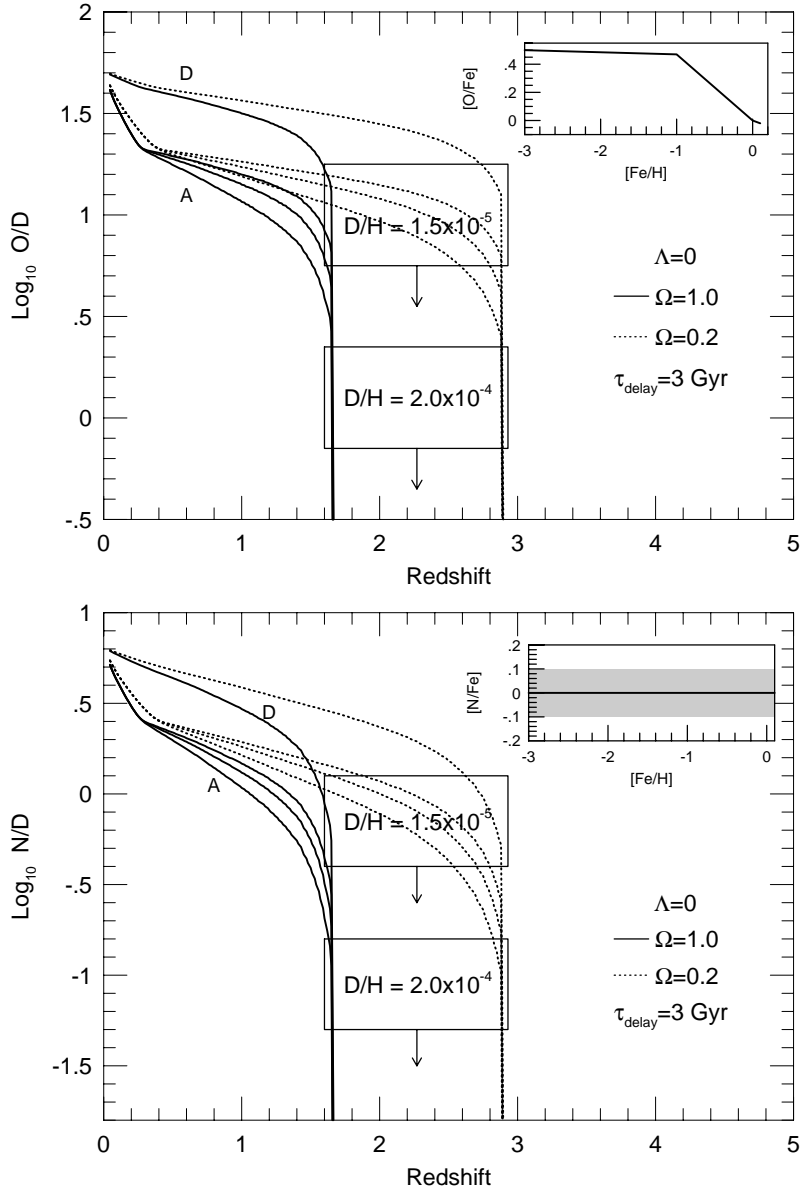

Fig. 3.— Predicted redshift histories of the (a) O/D and (b) N/D ratios. Inset plots show the general trends of [O/Fe] and [N/Fe] with [Fe/H], as measured from surveys of nearby dwarf stars. See text for transformation of the inset plots into the main plots. Curves labeled A–D have the same meanings as in Fig. 1a, with labels B and C omitted for clarity. Boxed regions are the predicted ranges for O/D for primordial D/H ratios of $1.5\times10^{-5}$ and $2 \times 10^{-4}$, as labeled, and they determine the initial coordinates from which the boxes are drawn. The range of O and N abundances, which determined the vertical dimension of each box, were deduced from Galactic halo star ratios in conjunction with the observed column densities of Si II and Fe II in QSOALS. Horizontal dimensions of each box were set by the cosmological models considered. Downward pointing arrows on each box indicate that these regions are upper limits, since Fe II and Si II lines were used in making the inferences and abundances of these species will be overestimated if any of the observed ions come from the H II regions. For a delay time of 1 Gyr instead of the 3 Gyr shown, the vertical part of the solid curve moves to z $\sim$ 4.4.

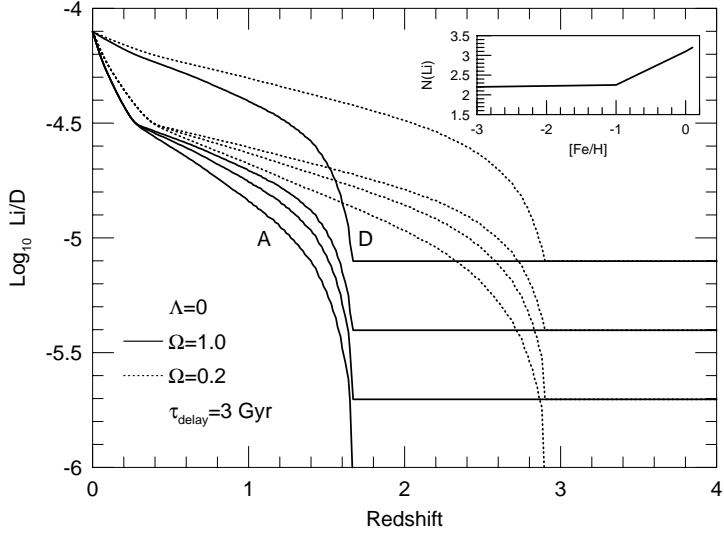

Fig. 4.— Redshift evolution of the lithium to deuterium ratio. Small inset plot shows a schematic of the upper envelope of lithium abundances deduced from surveys of solar vicinity dwarfs. Transformation of the small inset plot into the main plot follows the previous prescriptions. Curves labeled A–D have the same meanings as in Fig. 1a, with labels B and C omitted for clarity. It is unlikely that lithium will ever be seen in QSOALS, but implications of a positive detection are discussed in the text. For a delay time of 1 Gyr instead of the 3 Gyr shown, the vertical part of the solid curve moves to z $\sim$ 4.4.

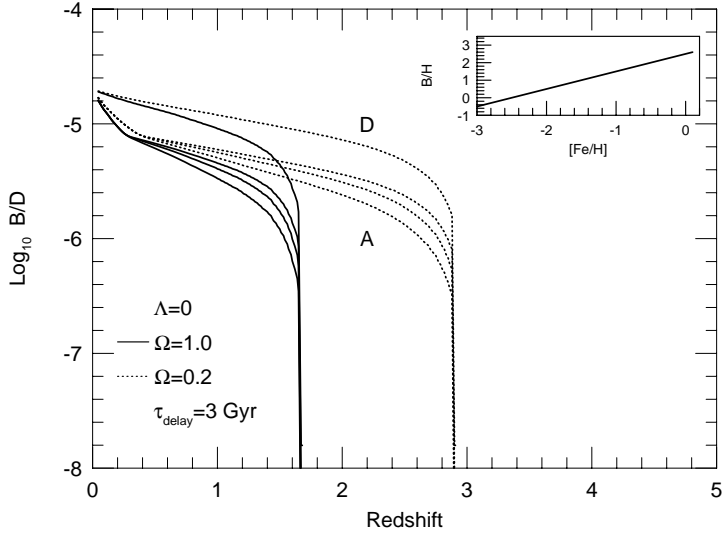

Fig. 5.— Redshift evolution of the boron to deuterium ratio. Small inset plot shows a schematic of the boron abundances inferred from surveys solar vicinity dwarfs. Transformation of the small inset plot into the main plot follows the previous mapping methods. Evolutions labeled A–D have the same meanings as in Fig 1a, with labels B and C omitted for clarity. Detection of the trace element boron will be difficult in QSOALS, but measurements may be able to assist in determining its chief nucleosynthetic origin site.

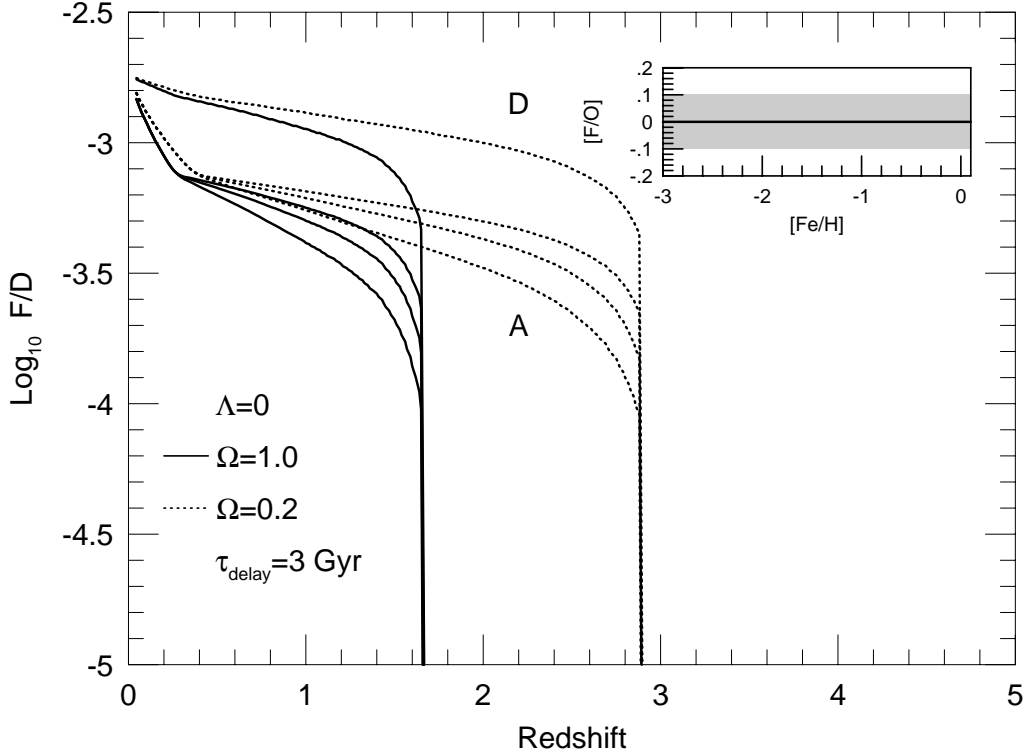

Fig. 6.— Fluorine to deuterium evolution with redshift. Small inset plot shows a schematic of the fluorine abundances inferred from the very few observations that exist of the hydrogen fluoride molecule in solar neighborhood stars. For $Z \lesssim -1.0$, the schematic evolution is guided by theoretical prejudices. Curves labeled A–D have the same meanings as in Fig. 1a, with labels B and C omitted for clarity. Fluorine will be very difficult to measure in QSOALS, but undisputed measurement of any fluorine abundance at a sufficiently large redshift ($z \gtrsim 1.5$) would constitute strong evidence for operation of the neutrino process in massive stars, and would influence views on the origin of elemental boron. For a delay time of 1 Gyr instead of the 3 Gyr shown, the vertical part of the solid curve moves to z $\sim$ 4.4.

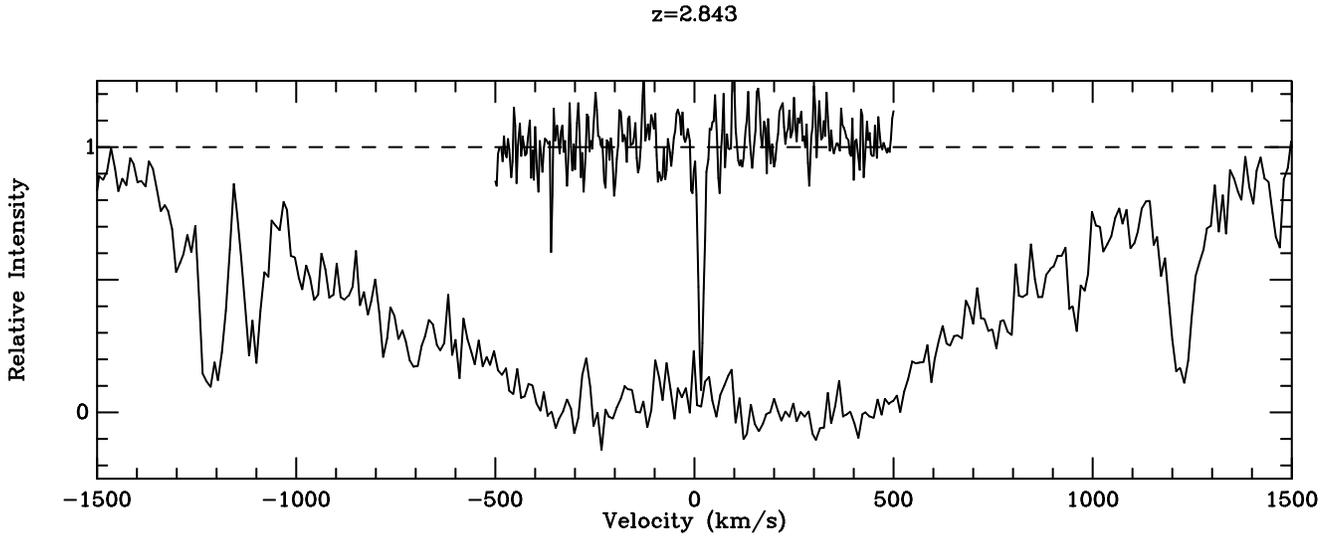

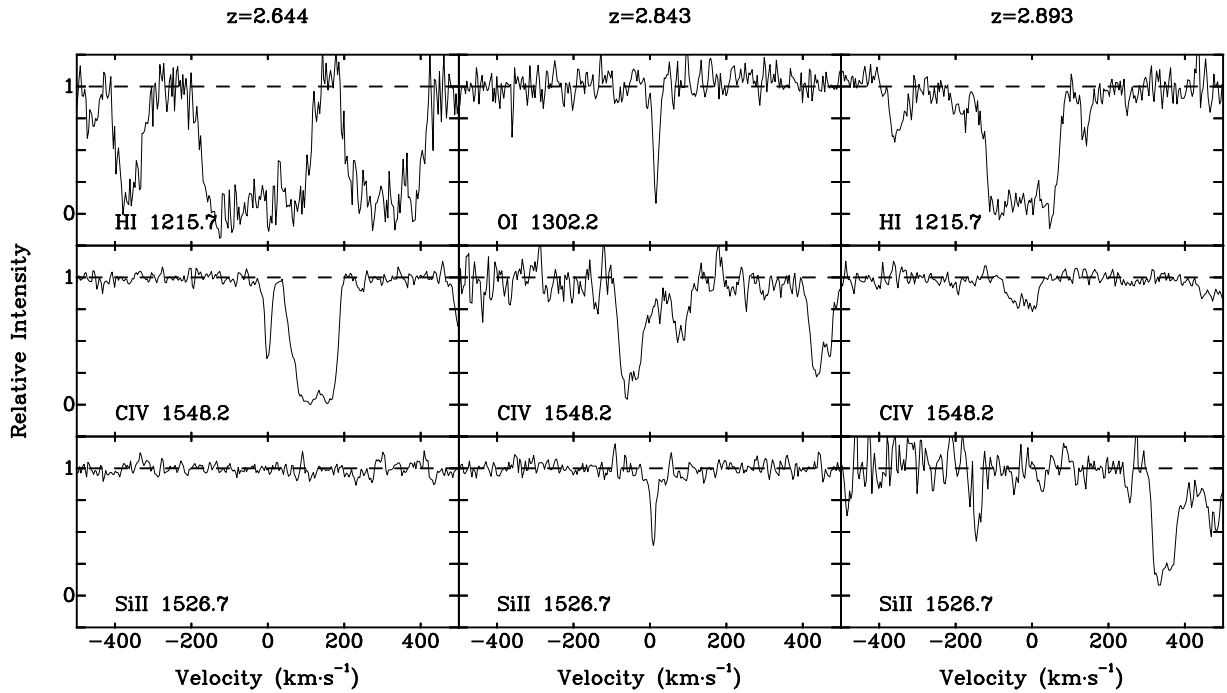

Fig. 7.— Relative intensity vs. H I, C IV, and Si II velocity for the QSOALS in HS 1946+7658 listed in Table 1. Note that the strong O I λ1302 line profile of the z=2.843 system is also plotted.

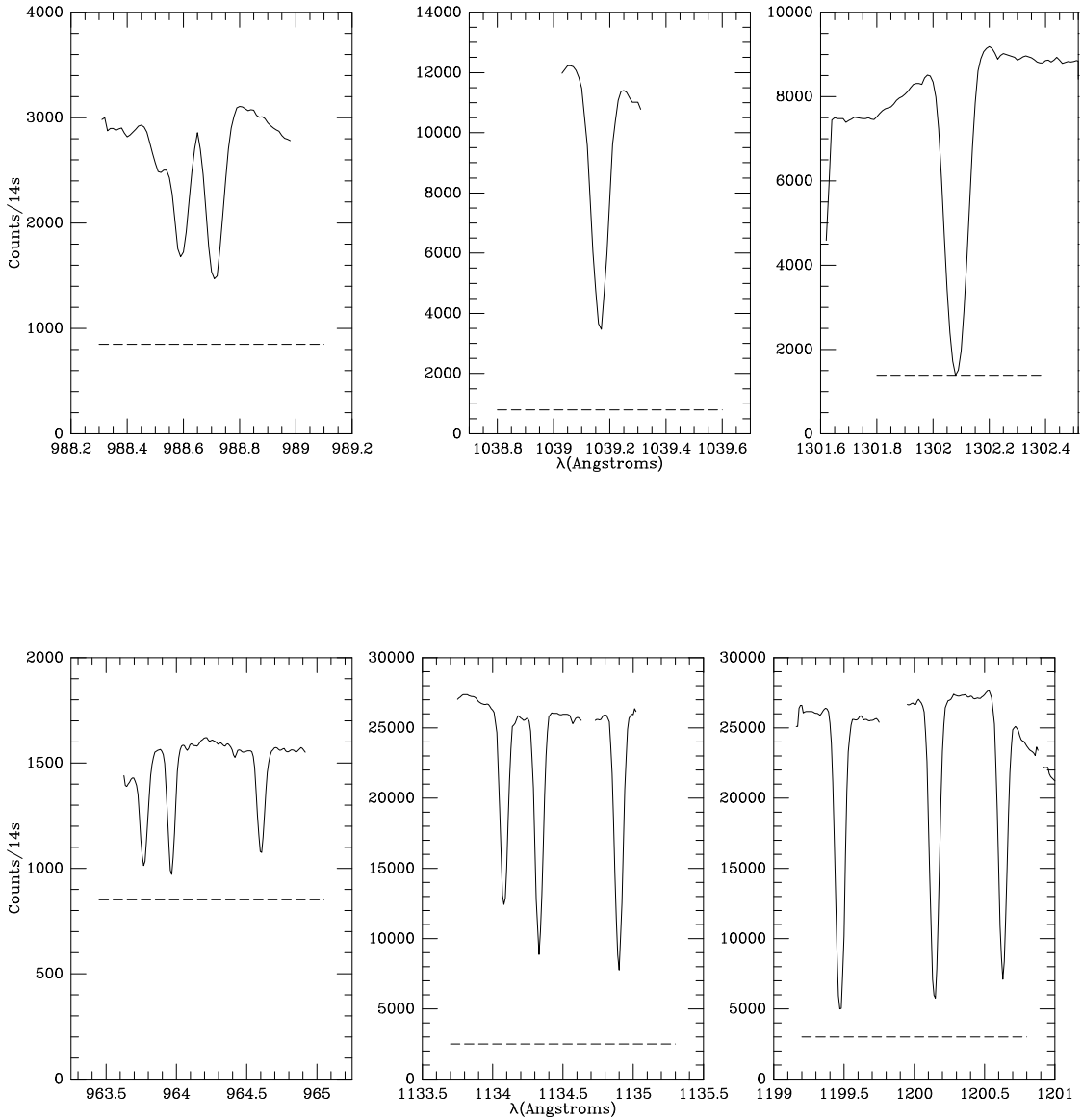

Fig. 8.— *Copernicus* observations of O I (top) and N I (bottom) transitions toward the star $\alpha$ Vir. These O I and N I features should be detectable in damped Ly$\alpha$ systems with H I column densities $\gtrsim 10^{20}$ cm$^{-2}$ and metallicities $\lesssim 0.01$ Z$_\odot$. The approximate background due to scattered light is shown as the dashed line.